\documentclass[11pt]{article}
\usepackage{amssymb}
\usepackage{amsmath}
\usepackage{graphics}
\usepackage{epsfig}
\usepackage{a4wide}
\usepackage{cite}
\usepackage{multirow}
\usepackage{color}
\usepackage{mathrsfs}

\DeclareGraphicsExtensions{.eps}
\begin{document}

\title{ NLO QCD and electroweak corrections to $WWW$ production at the LHC }
\author{
Shen Yong-Bai,$^1$ Zhang Ren-You,$^1$ Ma Wen-Gan,$^1$ Li Xiao-Zhou$,^1$ and Guo Lei$^2$ \\
{\small $^1$ Department of Modern Physics, University of Science and Technology of China (USTC),}  \\
{\small  Hefei 230026, Anhui, People's Republic of China} \\
{\small $^2$ Department of Physics, Chongqing University, Chongqing 401331, People's Republic of China}  }

\date{}
\maketitle
\vskip 15mm
\begin{abstract}
In this paper we calculate the next-to-leading order (NLO) QCD $[{\cal O}(\alpha_s \alpha^3)]$ and electroweak (EW) $[{\cal O}(\alpha^4)]$ corrections to the $WWW$ production at the LHC and deal with the subsequent leptonic decays from $W$ bosons by adopting an improved narrow-width approximation which takes into account the spin correlation and finite-width effects. The NLO QCD correction from the real jet radiation is discussed, which significantly enhances the production rate, particularly in the high-energy region. We also provide the integrated cross section for the $WWW$ production and various kinematic distributions of final products at the QCD+EW NLO. We find that the convergence of the perturbative QCD description can be improved by applying a hard jet veto in the event selection, but this jet veto would introduce a new source of theoretical uncertainty. The pure NLO QCD relative correction to the integrated cross section for the $W^+W^-W^+$ production at the $14~{\rm TeV}$ LHC is on the order of $30\%$ in the jet-veto event selection scheme with $p^{\, {\rm cut}}_{T,\, {\rm jet}} = 50~{\rm GeV}$, while the genuine NLO EW relative correction can reach about $15\%$ in the inclusive event selection scheme. Our numerical results show that both the NLO QCD and NLO EW corrections should be taken into consideration in precision predictions.
\end{abstract}

\vskip 35mm
{\large\bf PACS: 12.15.Lk, 12.38.Bx, 14.70.Fm }

\vfill \eject
\baselineskip=0.32in
%%%%%%%%%%%%%%%%%%%%%%%%%%%%%%%%%%%%%% Introduction %%%%%%%%%%%%%%%%%%%%%%%%%%%%%%%%%%%%%%%%%%%%%%%%%
\makeatletter      % '@' is now a normal "letter" for TeX
\@addtoreset{equation}{section}
\makeatother       % '@' is restored as a "non-letter" character for TeX
\vskip 5mm
\renewcommand{\theequation}{\arabic{section}.\arabic{equation}}
\renewcommand{\thesection}{\Roman{section}.}
\newcommand{\nb}{\nonumber}

%%%%%%%%%%%%%%%%%%%%%%%%%%%%%%%%%%%%%%%%%%%%%%%%%%%%%%%%%%%%%%%%%%%%%%%%%%%%%%%%%%
\vskip 5mm
\section{INTRODUCTION}
\par
With the discovery of the $126~ {\rm GeV}$ Higgs boson by the ATLAS and CMS Collaborations \cite{higgs-1,higgs-2}, one of the further goals of the Large Hadron Collider (LHC) is to test the Standard Model (SM). The precision test of the SM at the LHC requires accurate and reliable phenomenological predictions. The measurement of the gauge couplings is one of the most important experiments to study the gauge structure of the SM. The gauge couplings are usually investigated by the vector boson production processes and most of these processes at the LHC have been computed up to the QCD next-to-leading order (NLO) so far. The measurement at $14~{\rm TeV}$ will be possible in LHC Run 2 with higher luminosity, and the sensitivity to electroweak (EW) couplings increases with the reach in the high-energy tails of distributions. The NLO QCD prediction alone may not deliver a reliable estimate as expected. Theoretical precision predictions are, therefore, ahead of us to have a thorough interpretation of the data, which can be realized by taking into account the EW information. It is desired for the calculation of vector boson production with NLO QCD+EW accuracy.

\par
The triple vector boson production is sensitive to the triple and quartic gauge couplings (TGCs and QGCs) and, thus, related to the EW symmetry breaking mechanism \cite{TGC-1,QGC-2}. The measurements of triple gauge boson production at hadron colliders can provide rich information about the gauge boson self-interactions and play an important role in probing new physics beyond the SM. In order to improve the precision of the theoretical predictions, it is necessary to calculate the $VV^{\prime}V^{\prime\prime}~(V,V^{\prime},V^{\prime\prime} = W,~Z,~{\rm or}~\gamma)$ productions at hadron colliders up to the QCD+EW NLO including the subsequent vector boson decays, which are listed in the Les Houches 2015 precision SM wish list \cite{wishlist}. In previous works, all the triple gauge boson production processes at hadron colliders, i.e., $pp \rightarrow WWZ$, $ZZZ$, $WWW$, $WZZ$, $WW\gamma$, $ZZ\gamma$, $Z\gamma\gamma$, $\gamma\gamma\gamma$, $W\gamma\gamma$, and $WZ\gamma$, have been studied in the SM up to the QCD NLO \cite{WWZ, ZZZ, WWZ+ZZZ+WWW+WZZ, WWW-WZZ, WWr-ZZr, Zrr-rrr, Wrr, Wrr-decay, WZr}, while only the $pp \rightarrow WWZ$, $WZZ$, and $ZZZ$ processes have been complemented by the NLO EW corrections \cite{EWWWZ, EWWZZ, EWZZZ}, and for the latter two processes, the subsequent $W$- and $Z$-boson leptonic decays are included.

\par
In this work, we present the NLO QCD+EW corrected integrated cross section and some kinematic distributions for the $pp \rightarrow WWW + X$ production at the LHC, including the subsequent $W$-boson leptonic decays in an improved narrow-width approximation. The observables in the $WWW$ production could be sensitive to both the triple and quartic vector boson couplings and, thus, relevant to the study of these anomalous gauge couplings \cite{anomalous-coupling-1,anomalous-coupling-2}. The NLO QCD correction to the $WWW$ production at the LHC was provided in Refs.\cite{WWZ+ZZZ+WWW+WZZ, WWW-WZZ}. We hereby extend the calculation for the $WWW$ production to the QCD+EW NLO including subsequent $W$-boson leptonic decays to provide more accurate predictions. This paper is organized as follows: In Sec. II we provide our calculation strategy. The integrated cross section and various kinematic distributions are presented in Sec. III. Finally, a summary is given in Sec. IV.

\vskip 5mm
%%%%%%%%%%%%%%%%%%%%%%%%%%%%%%%%%%%%%%%%%%%%%%%%%%%%%%%%%%%%%%%%%%%%%%%%%%%%%%%%%%
\section{CALCULATIONS}
%%%%%%%%%%%%%%%%%%%%%%%%%%%%%%%%%%%%%%%%%%%%%%%%%%%%%%%%%%%%%%%%%%%%%%%%%%%%%%%%%%
\par
\subsection{General setup}
\par
In the LO and NLO QCD+EW calculations, we adopt the 't Hooft-Feynman gauge and only take into account the Cabibbo-Kobayashi-Maskawa (CKM) mixing between the first two quark generations. We set the masses of the first two generations of quarks to zero and adopt the four-flavor scheme in the initial-state parton convolution; therefore, there is no bottom-quark-induced subprocess, and the CKM matrix drops out in the flavor-summed closed quark loops. For all the partonic processes involved in the triple $W$-boson production at the QCD+EW NLO, i.e., $q\bar{q}^{\prime} \rightarrow WWW$, $q\bar{q}^{\prime} \rightarrow WWW + g$, $q\bar{q}^{\prime} \rightarrow WWW + \gamma$, $qg \rightarrow WWW + q^{\prime}$, and $q\gamma \rightarrow WWW + q^{\prime}$, the CKM matrix can factorize from all the amplitudes, and only one generic amplitude for each subprocess has to be evaluated in the parton distribution function (PDF) convolution \cite{EWWZZ,Wr}.

\par
We denote the ingredients of the calculation as below:
\begin{itemize}
\item $\sigma_{\rm LO}$: LO total cross section obtained by using LO PDFs
\item $\sigma_0$: LO total cross section obtained by using NLO PDFs
\item $\Delta \sigma_{\rm QCD}$: NLO QCD correction from the dynamic matrix element by using NLO PDFs
\item $\Delta \sigma_{\rm EW}$: NLO EW correction, which is the summation of $\Delta \sigma^{q\bar{q}}_{\rm EW}$ and $\Delta \sigma^{q\gamma}_{\rm EW}$, by using NLO PDFs. The superscripts $q\bar{q}$ and $q\gamma$ stand for the quark-antiquark and photon-induced subprocesses, respectively.
\end{itemize}
Unlike the QCD corrections from the quark-antiquark and gluon-induced channels, the EW corrections from the quark-antiquark and photon-induced subprocesses can be distinguished by their final-state products. With these definitions, the NLO QCD and NLO EW relative corrections to $pp \rightarrow WWW + X$ are given by
\begin{eqnarray}
\label{relativeQCDC}
&&
\delta_{\rm QCD}
=
\frac{\Delta \sigma_{\rm QCD} + \left(\sigma_0 - \sigma_{\rm LO}\right)}{\sigma_{\rm LO}}, \\
\label{relativeEWC}
&&
\delta_{\rm EW}
=
\delta^{q\bar{q}}_{\rm EW}
+
\delta^{q\gamma}_{\rm EW}
=
\frac{\Delta \sigma^{q\bar{q}}_{\rm EW}}{\sigma_{0}}
+
\frac{\Delta \sigma^{q\gamma}_{\rm EW}}{\sigma_{0}}.
\end{eqnarray}
In Eq.(\ref{relativeQCDC}), $\sigma_0 - \sigma_{\rm LO}$ is the NLO QCD contribution from the PDFs. The NLO EW correction is normalized by $\sigma_0$ in order to cancel the QCD contribution from the NLO PDFs. Therefore, $\delta_{\rm EW}$ defined in Eq.(\ref{relativeEWC}) is the genuine NLO EW relative correction which is practically independent of the PDF set.

\par
The pure NLO QCD corrected cross section $\sigma_{\rm QCD}$ can be expressed as
\begin{eqnarray}
\sigma_{\rm QCD} = \sigma_{\rm LO} \left( 1 + \delta_{\rm QCD} \right).
\end{eqnarray}
In order to include the potentially large contribution from the interplay between the EW and QCD corrections beyond the NLO, we calculate the combined NLO QCD+EW correction by using the naive product \cite{Denner-delta}
\begin{eqnarray}
\label{Xection-nlo}
\frac{\sigma_{\rm NLO}}{\sigma_{\rm LO}}
\equiv
1 + \delta_{\rm NLO}
=
\left( 1 + \delta_{\rm QCD} \right) \left( 1 + \delta^{q\bar{q}}_{\rm EW} +\delta^{q\gamma}_{\rm EW} \right).
\end{eqnarray}
This definition is particularly used in observables that receive extremely large QCD correction.

\par
At the parton level, the $W^-W^+W^-$ production is just the $CP$ conjugation of the $W^+W^-W^+$ production. Within the $CP$-conserved SM, the only difference between the $W^-W^+W^-$ and $W^+W^-W^+$ productions at the LHC is the initial-state parton convolution. Therefore, we describe the LO and NLO calculations only for the $pp \rightarrow W^+W^-W^+ + X$ process in the following. At the lowest order, the $W^+W^-W^+$ is produced via the quark-antiquark annihilation, i.e., $pp \rightarrow q \bar{q}^{\prime} \rightarrow W^+W^-W^+ + X~ (q = u, c,~ q^{\prime} = d, s)$. We can see clearly that the $WW\gamma$, $WWZ$ TGCs and $WWWW$ QGC are involved in some LO Feynman diagrams for the $q \bar{q}^{\prime} \rightarrow W^+W^-W^+$ partonic process.

\par
The final produced $W$ bosons are unstable particles. We will consider their leptonic decays in investigating the triple $W$-boson production at the LHC. The {\sc MadSpin} method \cite{madspin-2} is an improved narrow-width approximation based on the Frixione-Laenen-Motylinski-Webber (FLMW) \cite{madspin-1} approach which performs well in preserving the spin correlation and finite-width effects. In the FLMW approach, the off-shell effect is kept by smearing the mass of each resonance according to a Breit-Wigner distribution, and the spin correlation information is retained based on the acceptance-rejection method to generate the decay configuration. In this work we study only the triple physical $W$-boson production with subsequent leptonic decays, i.e., $pp \rightarrow WWW \rightarrow 3 \ell + 3 \nu + X$, while the $pp \rightarrow WH \rightarrow WW^*W \rightarrow 3 \ell + 3 \nu + X$ process is not included. We adopt the {\sc MadSpin} method in both the NLO QCD and NLO EW calculations to generate the final events in order to preserve the spin correlation and finite-width effects as far as possible. We first transform the differential cross sections for the $pp \rightarrow WWW + X$ process into Les Houches event files \cite{lhef-1,lhef-2}. Afterwards, we input these event files to the {\sc MadSpin} program, which is a part of the {\sc MadGraph5\_aMC@NLO} package \cite{mg5}, to generate the events after $W$-boson leptonic decays preserving both spin correlation and $W$-boson finite-width effects to a very good accuracy.

%%%%%%%%%%%%%%%%%%%%%%%%%%%%%%%%%%%%%%%%%%%%%%%%%%%%%%%%%%%%%%%%%%%%%%%%%%%%%%%%%%

\par
\subsection{Virtual corrections}
\par
The ultraviolet (UV) and infrared (IR) safety of physical observables requires the cancellation of singularities to all orders. In our calculation, the singularities are isolated by using the dimensional regularization scheme in $D=4-2\epsilon$ dimensions.  We adopt the on-mass-shell scheme to renormalize the masses and wave functions for the NLO QCD and EW corrections. The analytic expressions for the related EW renormalization constants and the unrenormalized EW self-energies can be found in Ref.\cite{W-physics}.

\par
The electric charge renormalization is introduced via the following relation,
\begin{eqnarray}
e^{\rm (B)} = \left( 1 + \delta Z_e \right) e ,
\end{eqnarray}
where $e^{\rm (B)}$ is the bare electric charge, and $\delta Z_e$ is the corresponding renormalization constant. By means of the on-shell condition for the $e^-$-$e^+$-$\gamma$ three-point Green function in the Thomson limit and the Ward identity, the electric charge renormalization constant in the $\alpha(0)$ scheme can be written as \cite{W-physics}
\begin{eqnarray}
\label{detZe0}
\delta Z^{\alpha(0)}_e
= -\frac{1}{2}\delta Z_{AA} - \frac{1}{2} \tan\theta_W \delta Z_{ZA}
=
\left[
\frac{1}{2}\frac{\partial \sum^{AA}_T(p^2) }{\partial p^2} - \tan\theta_W \frac{\sum^{AZ}_T(p^2)}{M^2_Z}
\right]_{p^2=0},
\end{eqnarray}
where $\theta_W$ is the weak mixing angle, and $\sum^{ab}_T(p^2)$ represents the transverse part of the unrenormalized self-energy of the $a \rightarrow b$ transition at four-momentum squared $p^2$. Equation (\ref{detZe0}) shows that the photon wave-function renormalization constant $\frac{1}{2}\delta Z_{AA}$ and, therefore, $\delta Z^{\alpha(0)}_e$, contains mass-singular terms $\ln(m_f^2/\mu^2)~ (f = e,\mu,\tau,u,d,c,s,b)$, but $\delta Z^{\alpha(0)}_e + \frac{1}{2}\delta Z_{AA}$ is free of these large logarithms. For a process with $l$ external photons and $n$ EW couplings in the lowest order amplitude, the wave-function renormalization of the $l$ external photons can only cancel the mass singularities from $l$ EW coupling counterterms; therefore, the full NLO EW correction would still contain residual mass singularities from the rest of the $n-l$ EW coupling counterterms if $n > l$. To obtain a more reliable perturbative prediction, we should use the running fine structure constant as input for $n-l$ EW vertices to absorb these unpleasant uncanceled large logarithms. As for the triple $W$-boson production considered in this paper, $l = 0$ and $n = 3$, we adopt the $G_{\mu}$ scheme for all three EW couplings at the LO. In the $G_\mu$ scheme, the fine structure constant is taken as
\begin{eqnarray}
\label{alphaGmu}
\alpha_{G_{\mu}} = \frac{\sqrt{2}G_{\mu} M^2_W}{\pi} \left(1-\frac{M^2_W}{M^2_Z}\right),
\end{eqnarray}
and the electric charge renormalization constant is correspondingly modified as
\begin{eqnarray}
\label{detZeGmu}
\delta Z_e^{G_\mu} = \delta Z_e^{\alpha(0)} - \frac{1}{2}\Delta r.
\end{eqnarray}
The subtraction term $\Delta r$ can be expressed as \cite{dtrform}
\begin{eqnarray}
\label{dtr}
\Delta r =
-\delta Z_{AA}
+
\Delta r^{\prime},
\end{eqnarray}
where
\begin{eqnarray}
\label{dtrprime}
\Delta r^{\prime}
=
- \cot^2\theta_W \left[ \frac{\sum^{ZZ}_T(M^2_Z)}{M^2_Z} - \frac{\sum^{WW}_T(M^2_W)}{M^2_W} \right]
+ \frac{\sum^{WW}_T(0)-\sum^{WW}_T(M^2_W)}{M^2_W} \nonumber \\
+ 2 \cot\theta_W \frac{\sum^{AZ}_T(0)}{M^2_Z}
+ \frac{\alpha(0)}{4\pi \sin^2\theta_W} \left[ 6 + \frac{7 -4 \sin^2\theta_W}{2 \sin^2\theta_W} \ln(\cos^2\theta_W) \right].  ~~~\,
\end{eqnarray}
Equations (\ref{detZeGmu})-(\ref{dtrprime}) clearly show that $\delta Z_e^{G_\mu}$ does not contain $\ln(m_f^2/\mu^2)$ because these logarithmic terms are absorbed by $\alpha_{G_{\mu}}$ \cite{W-physics,W-production}. For the additional EW couplings appearing at the EW NLO, we stick to employing the $\alpha(0)$ scheme. Then the LO cross section and the NLO EW correction are of the order of $\alpha_{G_{\mu}}^3$ and $\alpha_{G_{\mu}}^3 \alpha(0)$, respectively.

\par
After performing the renormalization procedure, the QCD/EW one-loop virtual correction to the $q \bar{q}^{\prime} \rightarrow W^+W^-W^+$ partonic process is UV finite but still contains soft and collinear IR singularities. The cancellation of IR singularities at QCD/EW NLO requires the gluon/photon bremsstrahlung. The soft IR singularity is canceled exactly by that in the real gluon/photon bremsstrahlung, while the collinear IR singularity is only partially canceled, and the remaining collinear IR singularity is absorbed by the collinear gluon/photon emission part of the related quark PDF QCD/EW counterterms. The analytic expressions for the quark PDF QCD and EW counterterms are provided in Ref.\cite{EWWZZ}.

\par
The Feynman amplitudes are created and simplified by using the modified {\sc FeynArts-3.7} \cite{feynarts} and {\sc FormCalc-7.3} \cite{formcalc}  packages. The one-loop amplitude is expressed as a linear combination of multipoint integrals. The five-point integrals are directly reduced to four-point integrals by using the Denner-Dittmaier method \cite{5-point-reduction}, and all the $N$-point ($N \leq 4$) tensor integrals are computed by means of the Passarino-Veltman reduction formalism \cite{PV-reduction}. In the calculation of four-point tensor integrals with rank $n > 3$, the numerical instability would occur at some phase-space region with small Gram determinant. We developed the codes for the loop calculation based on the {\sc LoopTools-2.8} package \cite{ff}, which can switch to the quadruple precision automatically in the region of
\begin{eqnarray}
 \frac{{\rm det}G_3}{(2 p_{\rm max}^2)^3} < \varepsilon,
\end{eqnarray}
where ${\rm det}G_3$ is the Gram determinant, and $p_{\rm max}^2$ is the maximum of the external four-momentum squared for a given four-point integral and $\varepsilon$ is set to $10^{-3}$ in this work. This algorithm to solve the unstable problem does not consume too much computer CPU time. In our test, we find that our improved program acts about 100 times faster than the pure quadruple precision codes in calculating the virtual corrections.\footnote{We used four i7-4790 3.60 GHz CPU cores to calculate the QCD virtual correction with $1 \times 10^{5}$ Monte Carlo samples by using pure quadruple precision and our developed codes, which consume 675 and 6.5 min, respectively.}

%%%%%%%%%%%%%%%%%%%%%%%%%%%%%%%%%%%%%%%%%%%%%%%%%%%%%%%%%%%%%%%%%%%%%%%%%%%%%%%%%%

\subsection{Real corrections}
\par
The QCD/EW real correction to the parent process $pp \rightarrow W^+ W^- W^+ + X$ originates from the gluon/photon bremsstrahlung and gluon-/photon-induced channels. We adopt the two cutoff phase-space slicing technique  \cite{tcpss} to isolate the IR singularities for the gluon/photon bremsstrahlung partonic process $q\bar{q}^{\prime} \rightarrow W^+W^-W^+ + g/\gamma$. Two cutoffs $\delta_{s}$ and $\delta_{c}$ are introduced to separate the phase space into soft ($E_{g/\gamma} \leqslant \delta_{s} \sqrt{\hat{s}}/2$), hard collinear ($E_{g/\gamma} > \delta_{s} \sqrt{\hat{s}}/2,~ \min\{ \hat{s}_{qg/\gamma},$ $\hat{s}_{\bar{q}^{\prime}g/\gamma}\} \leqslant \delta_{c}\hat{s}$), and hard noncollinear ($E_{g/\gamma} > \delta_{s} \sqrt{\hat{s}}/2$, $\min\{ \hat{s}_{qg/\gamma},~ \hat{s}_{\bar{q}^{\prime}g/\gamma}\} > \delta_{c}\hat{s}$) regions, where $\hat{s}_{ij} = (p_i + p_j)^2$, and $\sqrt{\hat{s}}$ is the colliding energy in the center-of-mass system. The soft and collinear IR singularities are located in the soft and hard collinear regions, respectively, while the phase-space integration over the hard noncollinear region is IR finite. Different from the gluon/photon bremsstrahlung, the gluon-/photon-induced partonic channel $qg/\gamma \rightarrow W^+W^-W^+ + q^{\prime}$ only contains collinear IR singularity, which can be canceled exactly by the collinear quark emission part of the related PDF counterterms. Therefore, the phase space is only separated into collinear ($\hat{s}_{q^{\prime}g/\gamma} \leqslant \delta_{c}\hat{s}$) and noncollinear ($\hat{s}_{q^{\prime}g/\gamma} > \delta_{c}\hat{s}$) regions for the isolation of the collinear IR singularity.

\par
The ${\cal O}(\alpha_s)$ QCD correction to the triple $W$-boson production at the LHC has already been investigated in Refs.\cite{WWZ+ZZZ+WWW+WZZ, WWW-WZZ}, but the intermediate Higgs boson exchange and $W$-boson leptonic decays were not taken into account in Ref.\cite{WWZ+ZZZ+WWW+WZZ}. To check the correctness of our calculation, we make a comparison with the results in Refs.\cite{WWZ+ZZZ+WWW+WZZ, WWW-WZZ} by using the same inputs and settings as in Ref.\cite{WWZ+ZZZ+WWW+WZZ}. From Table \ref{tab1} we see clearly that all these numerical results are in good agreement with each other within the Monte Carlo errors. However, we should mention that the contribution of the Feynman graphs with Higgs boson exchange amounts to about $3\%$ of the total cross section. We shall include all these Higgs boson exchange diagrams in the perturbative calculation.
%>>>>>>>>>>>>>>>>>>>>>>>>>>>>>>>>>>>>>>>>>>>>>>>>>>>>>>>>> Table 1
\begin{table}[htbp]
\begin{center}
\renewcommand\arraystretch{1.5}
\begin{tabular}{|c|ccc|ccc|}
\hline
\multirow{2}{*}{$\mu$}  & \multicolumn{3}{c|}{$\sigma_{\rm LO}~[fb]$} & \multicolumn{3}{c|}{$\sigma_{\rm QCD}~[fb]$} \\
\cline{2-7}
& Ref.\cite{WWZ+ZZZ+WWW+WZZ} & Ref.\cite{WWW-WZZ} & Ours & Ref.\cite{WWZ+ZZZ+WWW+WZZ} & Ref.\cite{WWW-WZZ} & Ours \\
\hline
\hline
$3M_Z/2$ & 82.7(5) & 82.7(1) & 82.62(3) & 153.2(6) & 152.5(3) & 152.44(9) \\
$3M_W$   & 82.5(5) & 82.8(1) & 82.74(3) & 146.2(6) & 145.2(3) & 145.17(6) \\
$6M_Z$   & 81.8(5) & 82.4(1) & 82.47(3) & 139.1(6) & 136.8(3) & 136.89(8) \\
\hline
\end{tabular}
\caption{
\label{tab1}
\small
Comparison between our results and the corresponding ones in Refs.\cite{WWW-WZZ, WWZ+ZZZ+WWW+WZZ} for $pp \rightarrow W^+W^-W^+ + X$. All input parameters and settings are taken from Ref.\cite{WWZ+ZZZ+WWW+WZZ}. }
\end{center}
\end{table}
%<<<<<<<<<<<<<<<<<<<<<<<<<<<<<<<<<<<<<<<<<<<<<<<<<<<<<<<<<

%%%%%%%%%%%%%%%%%%%%%%%%%%%%%%%%%%%%%%%%%%%%%%%%%%%%%%%%%%%%%%%%%%%%%%%%%%%%%%%%%%

\vskip 5mm
\section{RESULTS AND DISCUSSION}
The related SM input parameters are taken as \cite{PDG}
\begin{eqnarray}
M_W = 80.385~{\rm GeV},~~ M_Z = 91.1876~{\rm GeV},~~ M_t = 173.21~{\rm GeV},~~ M_H = 125.09~{\rm GeV}, \nonumber \\
G_F = 1.16638\times10^{-5}~{\rm GeV}^{-2},~~ \alpha(0) = 1/137.036,~~ \alpha_s(M_Z) = 0.119.~~~~~~~~~~~
\end{eqnarray}
All leptons and quarks except the top quark are treated as massless particles, and the CKM matrix elements are taken as
\begin{equation}
V_{\rm CKM} =
\left(
\begin{array}{ccc}
~~\,0.97425  & 0.22547 & 0\\
-0.22547 & 0.97425 & 0\\
~~\,0 & 0 & 1
\end{array}
\right).
\end{equation}
The factorization and renormalization scales are set to be equal, i.e., $\mu_F = \mu_R = \mu$, and the central scale is chosen as $\mu_0 = 3 M_W/2$. In the NLO calculation, we employ the NLO NNPDF2.3QED PDFs \cite{NNPDF} with $\overline{MS}$ and deep-inelastic scattering \cite{DIS-scheme} factorization schemes and set $\delta_s = 50 \times \delta_c = 10^{-3}$ and $ 10^{-4}$ for the QCD and EW corrections, respectively. The strong coupling constant is renormalized in the $\overline{MS}$ scheme with five active flavors, and its running is provided by the PDF set.

%%%%%%%%%%%%%%%%%%%%%%%%%%%%%%%%%%%%%%%%%%%%%%%%%%%%%%%%%%%%%%%%%%%%%%%%%%%%%%%%%%

\subsection{Integrated cross sections}
\label{Sec-IXection}
\par
The event selection scheme without any kinematic cuts on the final state is called the inclusive scheme. In the inclusive event selection scheme, the NLO QCD relative correction to $pp \rightarrow W^+W^-W^+ + X$ at the $14~ {\rm TeV}$ LHC is about $108\%$. The gluon-induced real correction and the QCD correction from quark-antiquark annihilation\footnote{In addition to $q\bar{q}^{\prime} \rightarrow WWW$, the real gluon and photon bremsstrahlungs $q\bar{q}^{\prime} \rightarrow WWW + g$ and $q\bar{q}^{\prime} \rightarrow WWW + \gamma$ are also classified as the quark-antiquark annihilation.} amount to $46\%$ and $54\%$ of the full NLO QCD correction, respectively. To improve the convergence of the perturbative QCD description, we may impose a tight jet veto on the final state to suppress the large QCD correction induced by the real jet radiation. In this paper, the exclusive event selection scheme with a jet transverse momentum cut of $p_{T,\, {\rm jet}} < p^{\, {\rm cut}}_{T,\, {\rm jet}} = 50~ {\rm GeV}$ is named the jet-veto scheme. In Table \ref{tab2} we provide the cross sections and NLO relative corrections for $pp \rightarrow W^+W^-W^{\pm} + X$ at the $8$ and $14~{\rm TeV}$ LHC in both inclusive and jet-veto event selection schemes. From this table, we can see that the photon-induced channels contribute a considerable amount of EW correction to the $WWW$ production in the inclusive event selection scheme. For the $WZZ$, $WWZ$, $W\gamma$, $WW$, and $WZ$ productions at the LHC \cite{WW-WZ,EWWWZ,Denner-delta,EWWZZ}, the NLO EW corrections from photon-induced channels are also sizable and can reach a few or a dozen percent of the corresponding total cross sections. In analogy to the QCD real jet radiation, the photon-induced EW correction can be heavily suppressed by applying a jet veto. For example, the photon-induced EW relative correction to the $W^+W^-W^+$ production at the $14~{\rm TeV}$ LHC is about $19\%$ in the inclusive event selection scheme but is reduced to about $2\%$ in the jet-veto event selection scheme. The EW correction from quark-antiquark annihilation is negative and independent of the jet veto. At the $14~ {\rm TeV}$ LHC, the full NLO EW relative correction to the $W^+W^-W^+$ production in the inclusive event selection scheme can reach about $15\%$.
%>>>>>>>>>>>>>>>>>>>>>>>>>>>>>>>>>>>>>>>>>>>>>>>>>>>>>>>>> Table 2
\begin{table}[htbp]
\begin{center}
\renewcommand\arraystretch{1.5}
\begin{tabular}{|c|c|c|c|c|}
\hline
\multirow{2}{*}{~~~~~~~~~~~~~~~}  & \multicolumn{2}{c|}{$\sqrt{S}=8~{\rm TeV}$} & \multicolumn{2}{c|}{$\sqrt{S}=14~{\rm TeV}$} \\
\cline{2-5}
& $W^+W^-W^+$ & $W^+W^-W^-$ & $W^+W^-W^+$ & $W^+W^-W^-$ \\
%& $pp \rightarrow W^+W^-W^+$ & $pp \rightarrow W^+W^-W^-$ & $pp \rightarrow W^+W^-W^+$ & $pp \rightarrow W^+W^-W^-$ \\
\hline
\hline
$\sigma_{\rm LO}~~$ $(fb)$         & $32.973(6)$  & $15.487(3)$   & $78.65(1)$   & $41.862(9)$   \\
$\sigma_{\rm QCD}^{\rm (I)}~(fb)$  & $61.29(2)$   & $30.998(9)$   & $163.20(3)$  & $92.58(4)$    \\
$\sigma_{\rm QCD}^{\rm (II)}~(fb)$ & $43.69(2)$   & $22.175(9)$   & $100.82(4)$  & $57.31(4)$    \\
$\sigma_{\rm NLO}^{\rm (I)}~(fb)$  & $67.49(4)$   & $35.03(3)$    & $187.04(9)$  & $108.62(7)$   \\
$\sigma_{\rm NLO}^{\rm (II)}~(fb)$ & $42.97(3)$   & $22.01(3)$    & $98.92(6)$   & $56.74(5)$    \\
\hline
$\delta_{\rm QCD}^{\rm (I)}~(\%)$   & $85.88$     & $100.16$      & $107.50$     & $121.16$      \\
$\delta_{\rm QCD}^{\rm (II)}~(\%)$  & $32.50$     & $43.18$       & $28.19$      & $36.90$       \\
$\delta_{\rm EW}^{\rm (I)}~~(\%)$   & $10.11$     & $13.01$       & $14.61$      & $17.33$       \\
$\delta_{\rm EW}^{\rm (II)}~~(\%)$  & $-1.64$~    & $-0.75$~      & $-1.88$~     & $-0.99$~       \\
\hline
$\delta_{\rm EW}^{q\bar{q}}~~(\%)$        & $-3.55$~    & $-3.18$~      & $-4.16$~     & $-3.72$~      \\
$\delta_{\rm EW}^{q\gamma \rm (I)}~(\%)$  & $13.66$     & $16.19$       & $18.77$      & $21.05$       \\
$\delta_{\rm EW}^{q\gamma \rm (II)}(\%)$  & $1.91$      & $2.43$        & $2.28$       & $2.73$        \\
\hline
\end{tabular}
\caption{
\label{tab2}
\small
LO, NLO QCD, and NLO QCD+EW corrected integrated cross sections and the corresponding relative corrections for $pp \rightarrow W^+W^-W^{\pm} + X$ at the $8$ and $14~{\rm TeV}$ LHC in the inclusive (I) and jet-veto (II) event selection schemes. }
\end{center}
\end{table}
%<<<<<<<<<<<<<<<<<<<<<<<<<<<<<<<<<<<<<<<<<<<<<<<<<<<<<<<<<

\par
The photon-induced correction is the leading component of the NLO EW correction to the $WWW$ production. The photon-induced PDF uncertainty, i.e., the PDF uncertainty from photon-induced channels, is given by \cite{PDF-uncertainty}
\begin{eqnarray}
\varepsilon_{\rm PDF}^{q\gamma}
=
\frac{1}{\sigma_{\rm NLO}}
\times
\left[
\frac{N}{N-1}
\left(
\langle \,
\sigma_{\rm EW}^{q\gamma \,\, 2} \,
\rangle
-
\langle \,
\sigma_{\rm EW}^{q\gamma} \,
\rangle^2
\right)
\right]^{1/2},
\end{eqnarray}
where the photon-induced correction $\sigma_{\rm EW}^{q\gamma}$ is a functional of PDFs,\footnote{Strictly speaking, the functional $\sigma_{\rm EW}^{q\gamma}$ depends only on quark and photon PDFs, i.e., $\sigma_{\rm EW}^{q\gamma} = \sigma_{\rm EW}^{q\gamma}[\Phi_{q|P}, \Phi_{\gamma|P}]$.} the expectation operator $\langle \, \ldots \, \rangle$ is defined as
\begin{eqnarray}
\langle \, {\cal F} \, \rangle
=
\frac{1}{N}\sum_{i = 1}^{N} {\cal F}[\Phi^{(i)}],
\end{eqnarray}
and $\Phi^{(i)}~ (i =1, ..., N)$ are the replicas of PDFs in the Monte Carlo ensemble. This PDF uncertainty is a main source of theoretical error, since the photon luminosity in NNPDF PDFs suffers from huge uncertainty especially at large Bjorken $x$. By adopting the NNPDF23\_nlo\_as\_0119\_qed set and taking $N = 100$, we obtain $\varepsilon_{\rm PDF}^{q\gamma} = 9.7\%$ and $2.0\%$ for the $W^+W^-W^+$ production at the $14~ {\rm TeV}$ LHC in the inclusive and jet-veto event selection schemes, respectively.\footnote{When the paper was completed, we became aware of the existence of a new PDF set, LUXqed \cite{LUX-PDF}, for dealing with photon PDF uncertainties. The photon-induced cross sections that we obtain using this PDF set are $\rm 9.625 \pm 0.142~ (\Delta PDF)~ fb$ and $\rm 1.343 \pm 0.019~ (\Delta PDF)~ fb$ for the inclusive and exclusive event selection schemes, respectively.} We can see that the NLO corrected inclusive cross section is plagued by the photon-induced PDF uncertainty, which is of comparable size to the photon-induced correction and can be reduced significantly after applying a jet veto.

\par
The factorization/renormalization scale dependence is another important source of theoretical uncertainty. The factorization scale $\mu_{\rm F}$ is involved in all perturbative orders via the PDF convolution, while the renormalization scale $\mu_{\rm R}$ occurs only at high orders via the renormalization procedure because the strong interaction is not involved in the $W^+W^-W^+$ production at the LO. In Table \ref{tab3} we present the LO, NLO QCD, and NLO QCD+EW corrected cross sections for the $W^+W^-W^+$ production at the $14~ {\rm TeV}$ LHC for some typical values of the factorization/renormalization scale. The factorization/renormalization scale uncertainty is defined as
\begin{eqnarray}
\varepsilon_{\rm scale}
=
\frac{1}{\sigma(\mu_0)}
\,
%\times
\max
\Big\{
\big( \sigma(\mu) - \sigma(\mu^{\prime}) \big)
\,\Big|\,
\mu, \, \mu^{\prime} \in \big[ \mu_0/4, \, 4\mu_0 \big]
\Big\}.
\end{eqnarray}
Then we obtain
\begin{eqnarray}
\label{scale-intxsection}
&& \varepsilon_{\rm scale}^{\rm LO} = 4.3\% ~~~~~~~~~~~~~~~~~~~~~~~~~~~~~~~~~~~{\rm (LO)} \nonumber \\
&& \varepsilon_{\rm scale}^{\rm QCD(I)} = 25\%,~~~~ \varepsilon_{\rm scale}^{\rm QCD(II)} = 1.1\% ~~~~~~\,{\rm (QCD~ NLO)} \nonumber \\
&& \varepsilon_{\rm scale}^{\rm NLO(I)} = 23\%,~~~~ \varepsilon_{\rm scale}^{\rm NLO(II)} = 0.9\% ~~~~~~~{\rm (QCD+EW~ NLO)}~~~~~
\end{eqnarray}
where the superscripts (I) and (II) stand for the inclusive and jet-veto event selection schemes, respectively. The scale uncertainty at the LO is much less than at the NLO in the inclusive event selection scheme. But the LO scale uncertainty underestimates the theoretical error from high order contributions, because the strong coupling constant is not involved in the LO matrix element. At the NLO, the EW correction is insensitive to the factorization/renormalization scale; the scale uncertainty mainly comes from the QCD real jet radiation and, therefore, can be reduced remarkably by applying a jet veto on the final state. However, the logarithmic dependence of the jet veto, i.e., $\ln(p^{\, {\rm cut} \, \, 2}_{T,\, {\rm jet}}/\mu^2)$, would induce an additional theoretical uncertainty in the jet-veto event selection scheme \cite{jt-uncertainty}. This theoretical uncertainty can be improved by the resummation of the jet-veto logarithms which falls outside the scope of this paper.
%>>>>>>>>>>>>>>>>>>>>>>>>>>>>>>>>>>>>>>>>>>>>>>>>>>>>>>>>> Table 3
\begin{table}[htbp]
\begin{center}
\renewcommand\arraystretch{1.5}
\begin{tabular}{|c|c|c|c|c|c|}
\hline
\multirow{2}{*}{$\mu$}	&  \multirow{2}{*}{$\sigma_{\rm LO}~ (fb)$}   &  \multicolumn{2}{|c|}{Inclusive scheme}	  &  \multicolumn{2}{|c|}{Jet-veto scheme} \\
\cline{3-6}
& & $\sigma_{\rm QCD}^{\rm (I)}~ (fb)$  &  $\sigma_{\rm NLO}^{\rm (I)}~ (fb)$  &  $\sigma_{\rm QCD}^{\rm (II)}~ (fb)$  &  $\sigma_{\rm NLO}^{\rm (II)}~ (fb)$  \\
\hline
\hline
$\mu_0/4$       &  $76.00(1)$    &  $189.16(3)$    &  $214.5(1)$     &  $101.89(3)$    &  $99.33(6)$  \\
\hline
$\mu_0/2$       &  $77.63(1)$    &  $174.47(3)$    &  $198.79(9)$    &  $101.17(3)$    &  $98.94(6)$  \\
\hline
$\mu_0$         &  $78.65(1)$    &  $163.20(3)$    &  $187.04(9)$    &  $100.82(4)$    &  $98.92(6)$  \\
\hline
$2\mu_0$        &  $79.20(1)$    &  $154.68(3)$    &  $178.43(8)$    &  $100.94(3)$    &  $99.32(6)$  \\
\hline
$4\mu_0$        &  $79.36(1)$    &  $147.98(3)$    &  $171.91(7)$    &  $101.16(3)$    &  $99.85(5)$  \\
\hline
\end{tabular}
\caption{
\label{tab3}
\small
Factorization/renormalization scale dependence of $\sigma_{\rm LO}$, $\sigma_{\rm QCD}$, and $\sigma_{\rm NLO}$ for the $W^+W^-W^+$ production at the $14~{\rm TeV}$ LHC in the inclusive (I) and jet-veto (II) event selection schemes.}
\end{center}
\end{table}
%<<<<<<<<<<<<<<<<<<<<<<<<<<<<<<<<<<<<<<<<<<<<<<<<<<<<<<<<<

\subsection{Kinematic distributions}
\par
In this subsection, we present the LO, NLO QCD, and NLO QCD+EW corrected kinematic distributions of final $W$ bosons and their leptonic decay products for the $W^+W^-W^+$ production at the $14~ {\rm TeV}$ LHC in both inclusive and jet-veto event selection schemes. To demonstrate the NLO EW correction more clearly, the EW relative corrections from the quark-antiquark and photon-induced channels ($\delta_{\rm EW}^{q\bar{q}}$ and $\delta_{\rm EW}^{q\gamma}$) are provided separately.

\par
\begin{figure}[htbp]
\begin{center}
\includegraphics[width=0.49\textwidth]{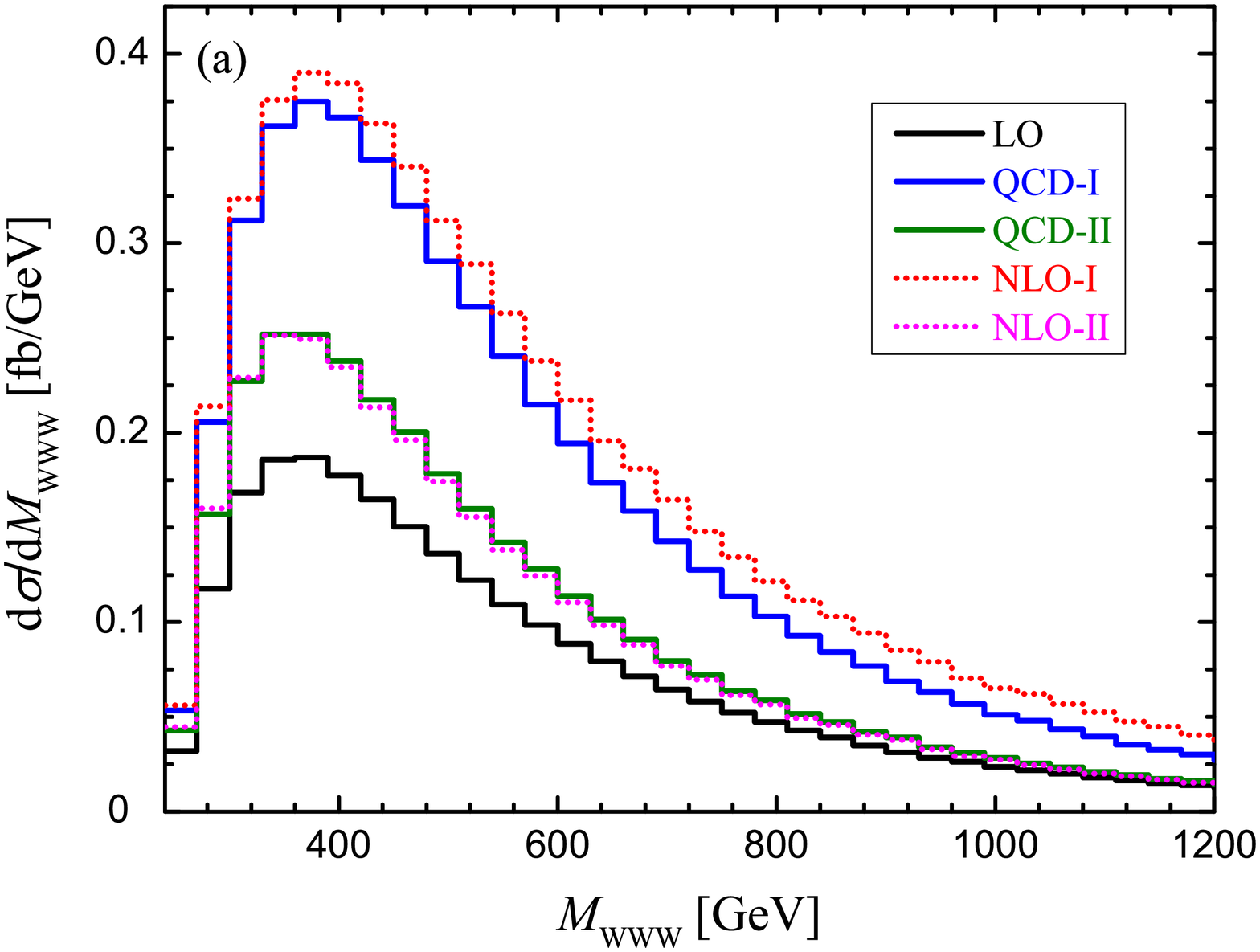}
\includegraphics[width=0.49\textwidth]{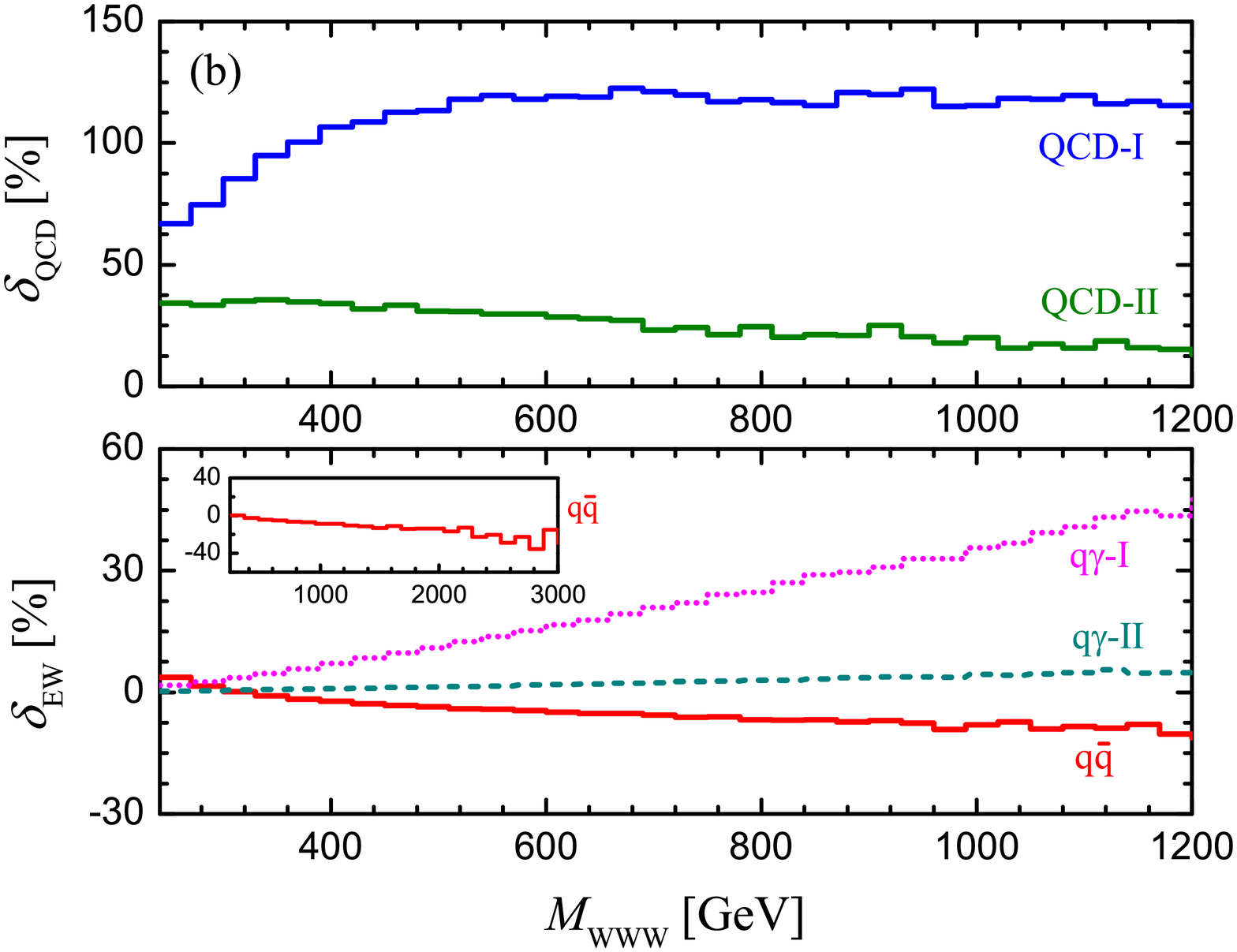}
\caption{
\label{fig-Mwww}
\small
(a) $W^+W^-W^+$ invariant mass distributions and (b) the corresponding NLO QCD and EW relative corrections for $pp \to W^+W^-W^+ + X$ at the 14 TeV LHC.
 }
\end{center}
\end{figure}

The LO, NLO QCD, and NLO QCD+EW corrected invariant mass distributions of the $W^+W^-W^+$ system are depicted in Fig.\ref{fig-Mwww}(a), and the corresponding QCD and EW relative corrections are provided in Fig.\ref{fig-Mwww}(b). The $W^+W^-W^+$ invariant mass distributions reach their maxima in the vicinity of $M_{WWW} \sim 360~ {\rm GeV}$ in both inclusive and jet-veto event selection schemes. The NLO QCD correction enhances the LO $W^+W^-W^+$ invariant mass distribution significantly, and the QCD relative correction exceeds $100\%$ when $M_{WWW} > 360~ {\rm GeV}$, in the inclusive event selection scheme. After applying the jet veto, the QCD relative correction is suppressed below $35\%$ and decreases slowly with the increment of $M_{WWW}$ in the plotted region. Compared to the jet-veto event selection scheme, the QCD relative correction in the inclusive event selection scheme increases rapidly in the region of $M_{WWW} < 360~ {\rm GeV}$. We may conclude that the real jet radiation is the dominant mechanism of the NLO QCD correction, even of comparable size to the LO prediction. To improve the convergence of the perturbative QCD description, we adopt the jet-veto scheme introduced in Sec. \ref{Sec-IXection}. In the jet-veto event selection scheme, the QCD real jet emission correction, especially the contribution from the gluon-induced channels, is heavily suppressed due to the jet veto, and, therefore, a more moderate and reliable NLO QCD correction can be obtained. From the lower panel of Fig.\ref{fig-Mwww}(b), we see that the EW relative correction from quark-antiquark annihilation decreases with the increment of $M_{WWW}$ and becomes negative when $M_{WWW} > 330~ {\rm GeV}$. It is sizable in the high invariant mass region and can exceed $-7\%$ when $M_{WWW} > 900~ {\rm GeV}$. Particularly, as the increment of $M_{WWW}$ to $3~ {\rm TeV}$, the Sudakov EW logarithms become large, and the relative EW correction to the $pp \rightarrow q\bar{q} \rightarrow WWW+X$ process is negative and of the order of tens percent. On the contrary, the photon-induced EW relative correction is positive and increases with the increment of $M_{WWW}$ in both inclusive and jet-veto event selection schemes. Analogous to the QCD real jet radiation, the photon-induced contribution (i.e., the EW real jet radiation) with a hard jet is treated as a $2 \rightarrow 4$ process and, therefore, subjected to the same cuts as the full QCD case. At $M_{WWW} \sim 1200~ {\rm GeV}$, $\delta_{\rm EW}^{q\bar{q}} \sim -10\%$, $\delta_{\rm EW}^{q\gamma \rm (I)} \sim 45\%$, and $\delta_{\rm EW}^{q\gamma \rm (II)} \sim 5\%$. It shows that the NLO EW correction is dominated by the hard jet radiation\footnote{The EW correction from the hard jet radiation is given by $\Delta_{\rm EW}^{\rm 1-jet} = \Delta_{\rm EW}^{q\gamma \rm (I)} - \Delta_{\rm EW}^{q\gamma \rm (II)}$.} in the high invariant mass region in the inclusive event selection scheme. In the jet-veto event selection scheme, the photon-induced correction is reduced since the hard jet radiation is excluded by the jet veto, and then the quark-antiquark annihilation becomes dominant in the high invariant mass region.
%--------------------------------------------------------->>>FIG
%---------------------------------------------------------<<<FIG

\par
\begin{figure}[htbp]
\begin{center}
\includegraphics[width=0.49\textwidth]{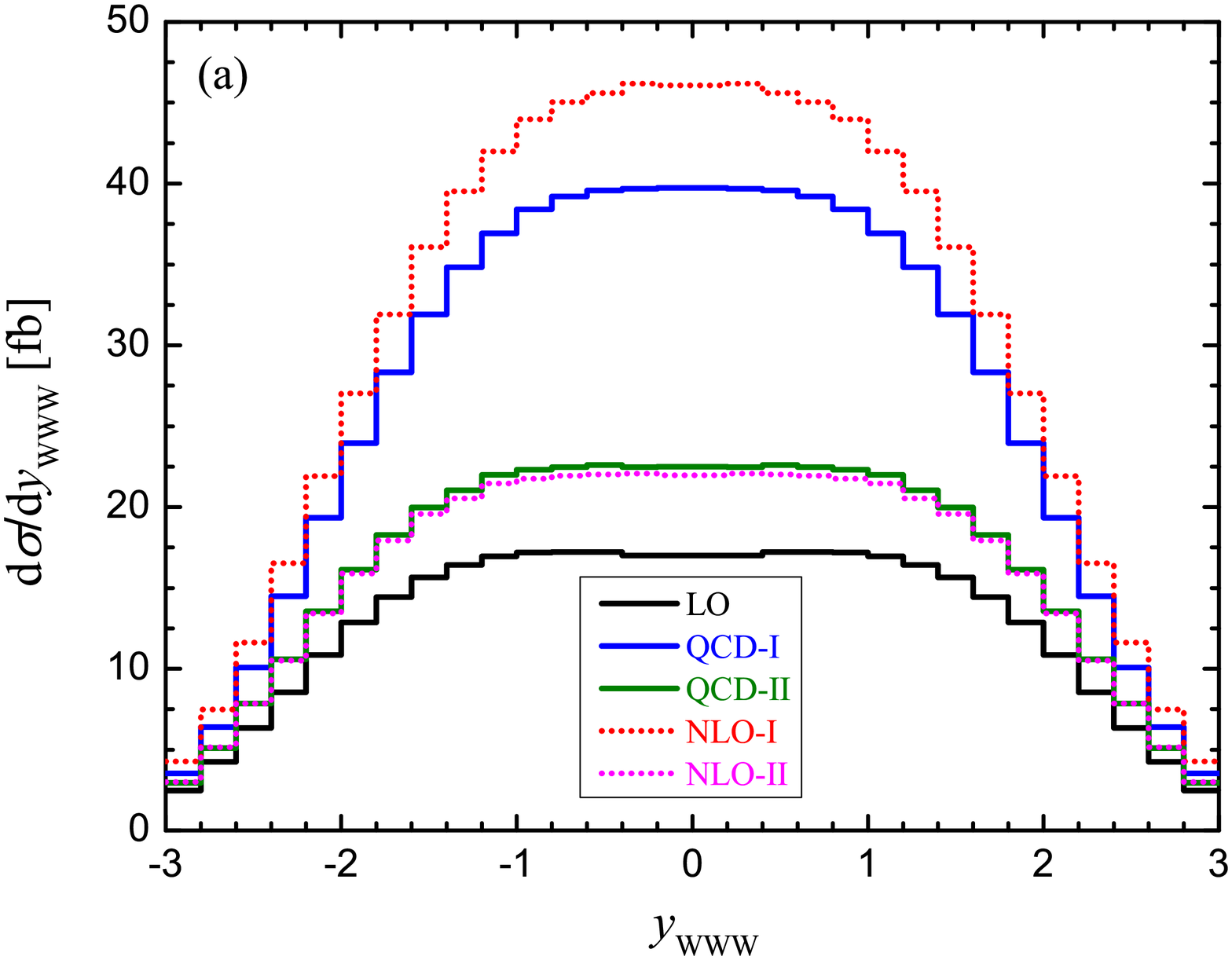}
\includegraphics[width=0.49\textwidth]{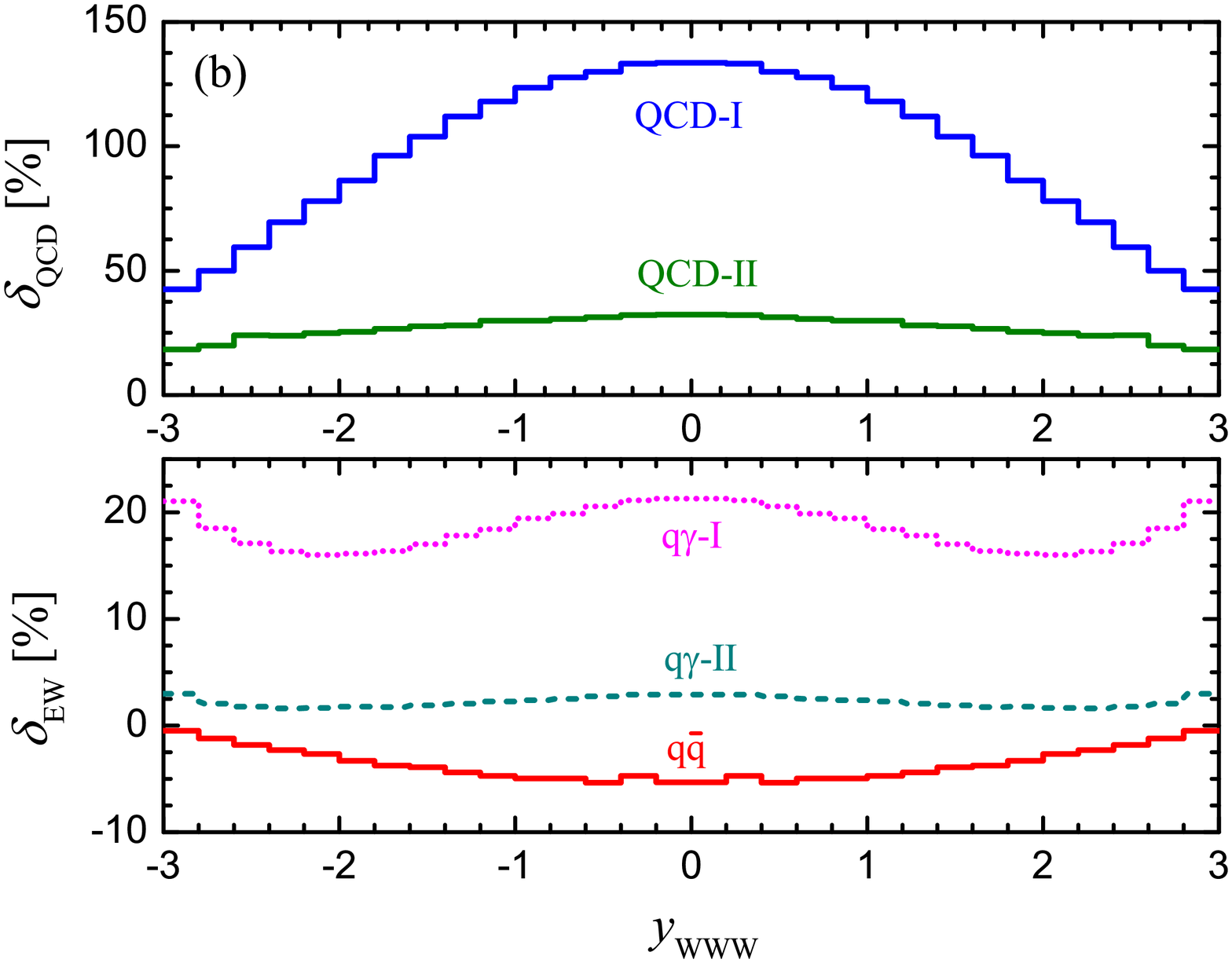}
\caption{
\label{fig-Ywww}
\small
(a) $W^+W^-W^+$ rapidity distributions and (b) the corresponding NLO QCD and EW relative corrections for $pp \to W^+W^-W^+ + X$ at the 14 TeV LHC.
 }
\end{center}
\end{figure}

The LO, NLO QCD, and NLO QCD+EW corrected rapidity distributions of the $W^+W^-W^+$ system and the corresponding QCD and EW relative corrections are provided in Figs.\ref{fig-Ywww}(a) and \ref{fig-Ywww}(b), respectively. The QCD relative correction to the $W^+W^-W^+$ rapidity distribution decreases rapidly from the order of $135\%$ to about $40\%$ with the increment of $|y_{WWW}|$ from $0$ to $3$ in the inclusive event selection scheme and varies from $30\%$ to $20\%$ correspondingly in the jet-veto event selection scheme. The $W^+W^-W^+$ events with a hard jet tend to be produced centrally, i.e., $y_{WWW} \rightarrow 0$, because the transverse momentum of $W^+W^-W^+$ is sufficiently large ($p_{T,\, WWW} > 50~ {\rm GeV}$) due to transverse momentum conservation. These $W^+W^-W^+ + {\rm jet}$ events will be completely excluded by adopting the jet-veto event selection scheme. As we expected, the QCD relative correction in the jet-veto event selection scheme is heavily reduced compared to the inclusive event selection scheme, particularly in the central rapidity region. The EW relative correction from quark-antiquark annihilation does not depend on the jet veto and increases slowly from $-5\%$ to $0$ as $|y_{WWW}|$ increases from $0$ to $3$. The photon-induced EW relative correction in the inclusive event selection scheme is quantitatively larger, varying in the range of $[ 15\% , \, 20\% ]$ approximately as $y_{WWW} \in [-3 , \, 3]$. After applying the jet veto, the hard jet radiation is excluded, and the photon-induced EW relative correction is steady at about $2\%$ in the whole plotted range. We can see that the quark-antiquark and photon-induced EW relative corrections are negative and positive, respectively, and are only a few percent in the jet-veto event selection scheme. Consequently, the full NLO EW relative correction in the jet-veto event selection scheme is negative in the region of $|y_{WWW}| < 2.6$, and its absolute value is less than $3\%$ in the whole plotted region. In Fig.\ref{fig-Ywww}(a), the two histograms labeled ``QCD-II" and ``NLO-II" substantially coincide with each other. It implies that the NLO EW correction modifies the $W^+W^-W^+$ rapidity distribution slightly in the jet-veto event selection scheme.
%--------------------------------------------------------->>>FIG
%---------------------------------------------------------<<<FIG

\par
\begin{figure}[htbp]
\begin{center}
\includegraphics[width=0.49\textwidth]{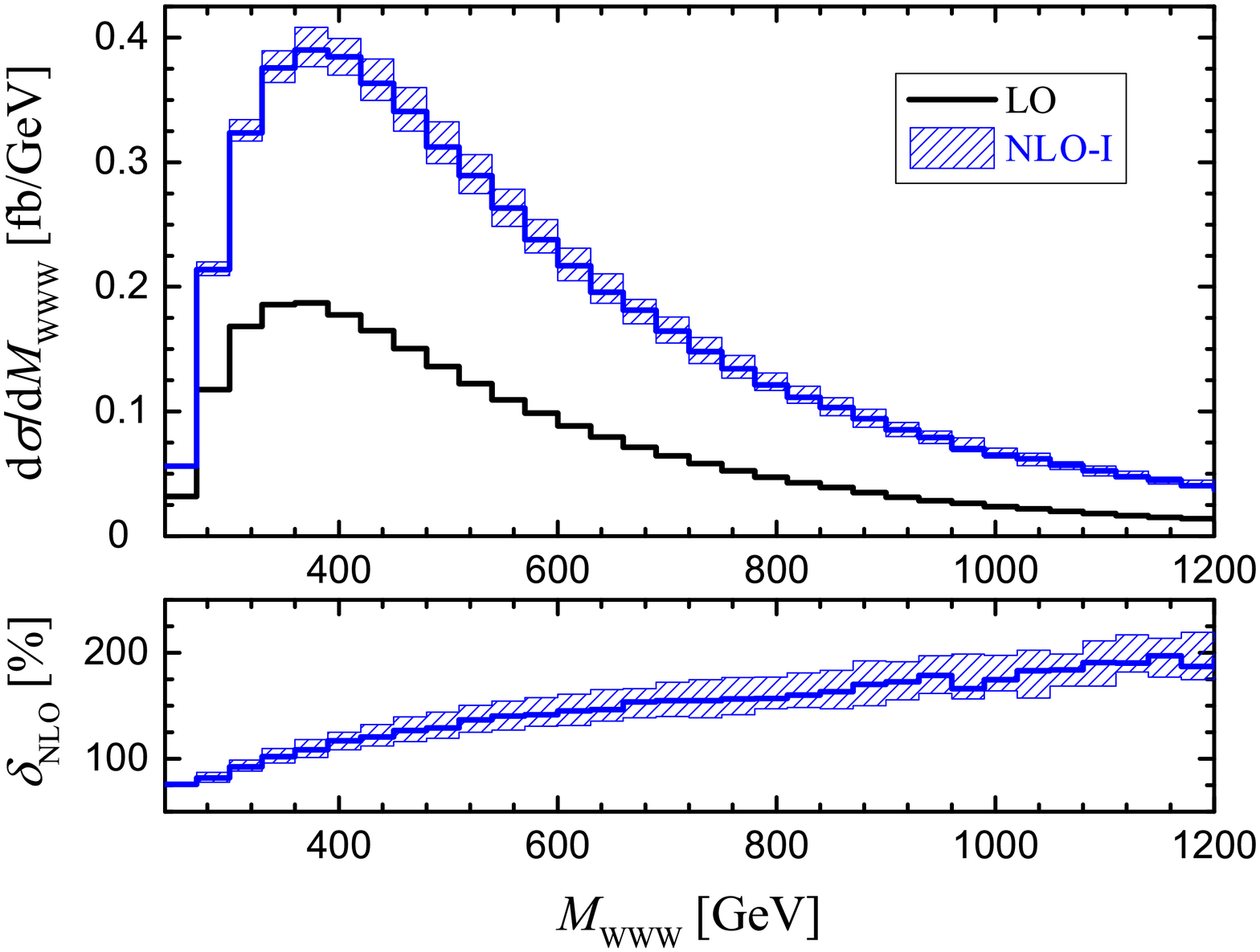}
\caption{
\label{fig-scale}
\small
NLO QCD+EW corrected inclusive $W^+W^-W^+$ invariant mass distribution and the corresponding QCD+EW relative correction for $pp \rightarrow W^+W^-W^+ + X$ at the $14~{\rm TeV}$ LHC with $\mu$ varying in the range of $[\mu_0/2 ,\, 2 \mu_0]$.
 }
\end{center}
\end{figure}

In order to display the scale uncertainty of differential cross sections, we plot the NLO QCD+EW corrected $W^+W^-W^+$ invariant mass distribution and the corresponding QCD+EW relative correction in the inclusive event selection scheme in Fig.\ref{fig-scale}, where the blue thick lines are the central predictions, and the blue bands reflect the uncertainty from the scale variation in the range of $[\mu_0/2 ,\, 2\mu_0]$. The LO $M_{WWW}$ distribution at the central scale (black thick line) is also provided in this figure only for comparison. The scale uncertainty of the NLO QCD+EW corrected inclusive differential cross section with respect to kinematic variable $x$ can be estimated by
\begin{eqnarray}
\varepsilon_{{\rm scale}}^{\rm NLO(I)}(x)
\,=\,
\frac{\left[d \sigma_{\rm NLO}^{\rm (I)}\Big/d x\right]_{\rm BW}}
{\left[d \sigma_{\rm NLO}^{\rm (I)}\Big/d x\right]_{\mu_0}~}
\, \sim \,
\frac{1}{\left[1 + \delta_{{\rm NLO}}^{\rm (I)}(x)\right]_{\mu_0}}
\times
\left[\delta_{{\rm NLO}}^{\rm (I)}(x)\right]_{\rm BW},
\end{eqnarray}
where the subscripts ${\rm BW}$ and $\mu_0$ stand for the band width and central value, respectively. Therefore, the $W^+W^-W^+$ invariant mass dependence of the scale uncertainty can be qualitatively or semiquantitatively described by the band in the nether panel of Fig.\ref{fig-scale}. It shows that the scale uncertainty increases with the increment of $M_{WWW}$, particularly in the low $M_{WWW}$ region, and tends to be independent of $M_{WWW}$ in the high $M_{WWW}$ region.
%--------------------------------------------------------->>>FIG
%---------------------------------------------------------<<<FIG

\par
Now we turn to the $W^+W^-W^+$ production with subsequent leptonic decays at the LHC, i.e., $pp \rightarrow W^+W^-W^+ \rightarrow 3 \ell + 3 \nu + X$, where the three charged leptons in the final state are $e$, $\mu$, and $\tau$, respectively. The branching ratios for $W$-boson leptonic decays and the total decay width of $\Gamma_{\rm total}^W = 2.045~ {\rm GeV}$ in the fixed-width scheme are obtained by using the {\sc MadSpin} program. The final-state charged leptons are required to have
\begin{eqnarray}
\label{bscuts}
p_{T,\, \ell} > 10~{\rm GeV} {\rm ~~~~~and~~~~~} |y_{\ell}| < 2.5.
\end{eqnarray}
The three identified leptons $\ell_1$, $\ell_2$, and $\ell_3$ are called leading, next-to-leading and next-to-next-to-leading leptons, respectively, according to their transverse momentum in decreasing order, i.e., $p_{T,\, \ell_1} > p_{T,\, \ell_2} > p_{T,\, \ell_3}$.

\par
\begin{figure}[htbp]
\begin{center}
\includegraphics[width=0.49\textwidth]{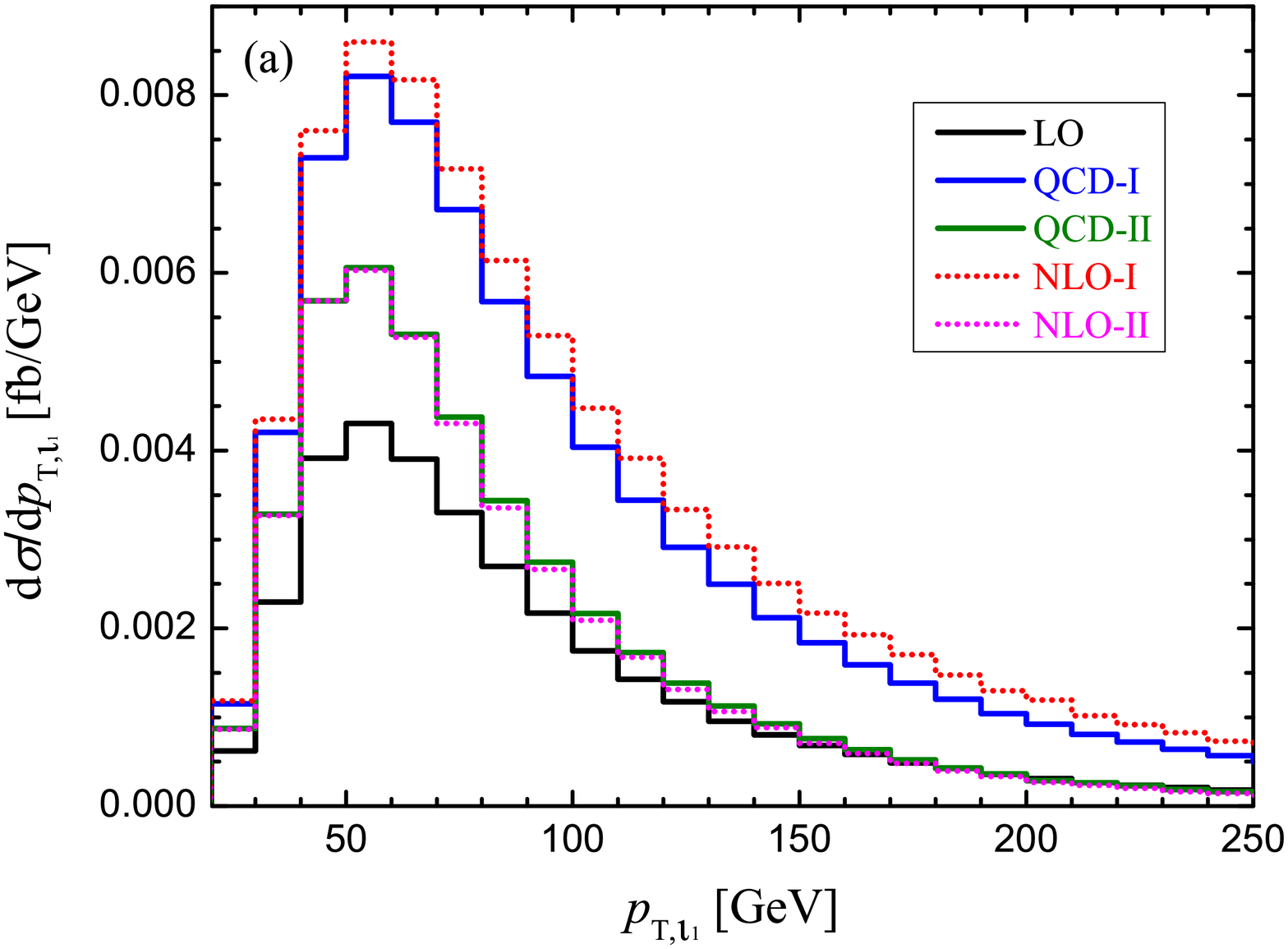}
\includegraphics[width=0.49\textwidth]{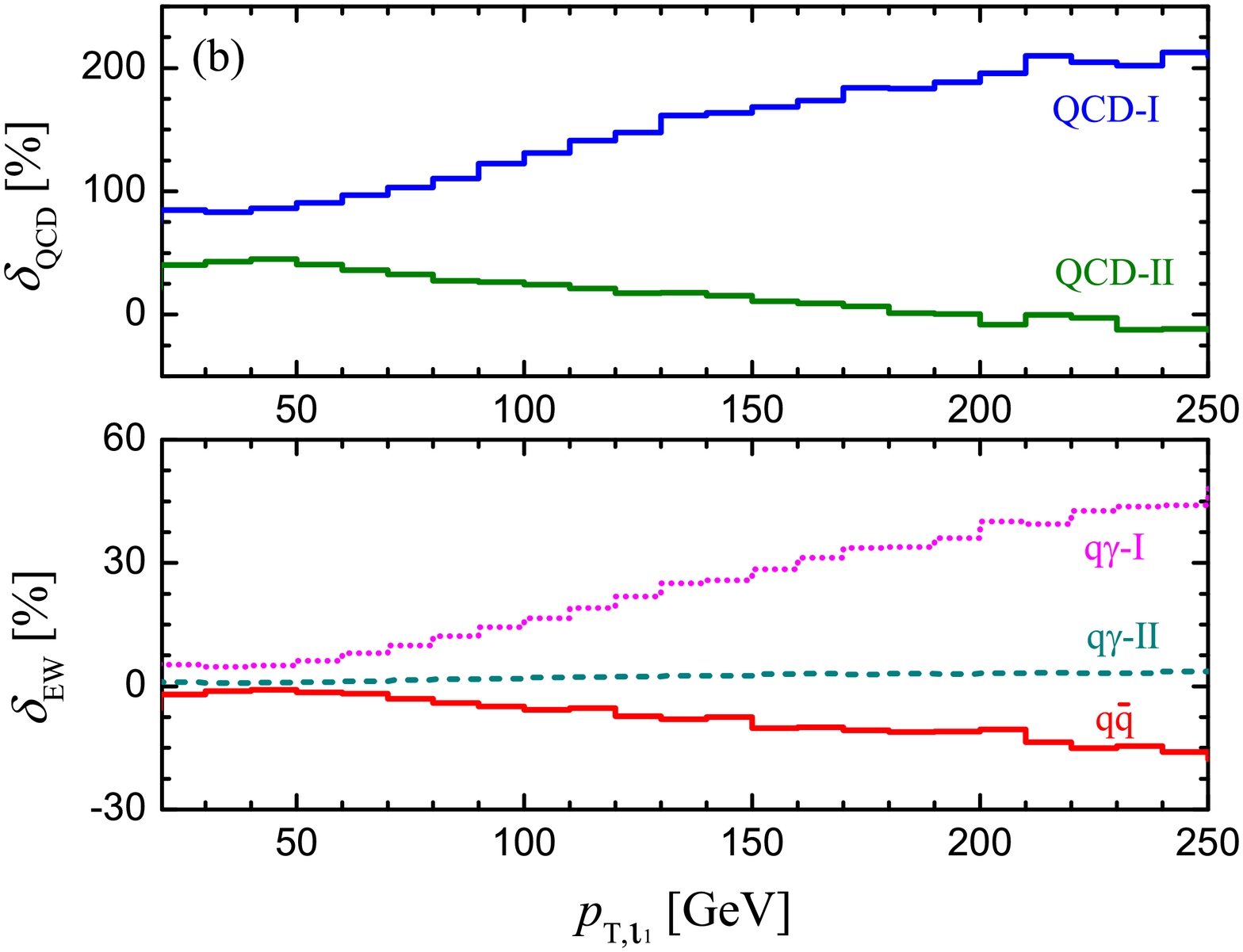}
\caption{
\label{fig-PTLl}
\small
(a) Transverse momentum distributions of the leading lepton and (b) the corresponding NLO QCD and EW relative corrections for $pp \to W^+W^-W^+ \to 3\ell + 3\nu + X$ at the 14 TeV LHC.
 }
\end{center}
\end{figure}

The LO, NLO QCD, and NLO QCD+EW corrected transverse momentum distributions of the leading lepton for $pp \rightarrow W^+W^-W^+ \rightarrow 3 \ell + 3 \nu + X$ at the $14~ {\rm TeV}$ LHC are shown in Fig.\ref{fig-PTLl}(a). The QCD and EW relative corrections are plotted in Fig.\ref{fig-PTLl}(b) correspondingly. All these lepton transverse momentum distributions peak at $p_{T,\, \ell_1} \sim 50~ {\rm GeV}$. In the inclusive event selection scheme, the QCD relative correction varies from $80\%$ to $210\%$ in the range of $20~ {\rm GeV} < p_{T,\, \ell_1} < 250~ {\rm GeV}$. The large QCD correction in the high $p_{T,\, \ell_1}$ region and $p_{T,\, \ell_1}$ dependence of the QCD relative correction can be traced to the kinematics of the gluon-induced channels \cite{WWW-WZZ,EWWWZ}. By contrast, the QCD relative correction in the jet-veto event selection scheme ranges from $-10\%$ to $45\%$ in the same $p_{T,\, \ell_1}$ range. It clearly shows that the jet veto significantly suppresses the QCD contribution from the real jet radiation and leads to a fairly moderate QCD relative correction. As shown in the lower panel of Fig.\ref{fig-PTLl}(b), the EW relative correction from quark-antiquark annihilation is negative and decreases from about $0$ to $-15\%$ with the increment of $p_{T,\, \ell_1}$ from $50$ to $250~ {\rm GeV}$, while the photon-induced EW relative correction in the inclusive event selection scheme is positive and increases correspondingly from $5\%$ to $45\%$. It shows that the EW correction in the inclusive event selection scheme is considerable, particularly in high transverse momentum region. However, by adopting the jet-veto event selection scheme, the photon-induced EW relative correction is heavily reduced and ranges only between $0$ and $5\%$ in the plotted $p_{T, \, \ell_1}$ region. The transverse momentum distributions of the next-to-leading and next-to-next-to-leading leptons ($\ell_2$ and $\ell_3$) are quite similar to the leading lepton $\ell_1$, and, therefore, are not discussed in this paper.
%--------------------------------------------------------->>>FIG
%---------------------------------------------------------<<<FIG

\par
To describe the orientation of the final produced leading lepton in more detail, we, respectively, demonstrate the distributions of the following two kinematic variables in this paper: (1) $\theta_{\ell_1\ell_2}$, the separation angle between the leading and next-to-leading leptons, and (2) $y_{\ell_1}$, the rapidity of the leading lepton.

\par
\begin{figure}[htbp]
\begin{center}
\includegraphics[width=0.49\textwidth]{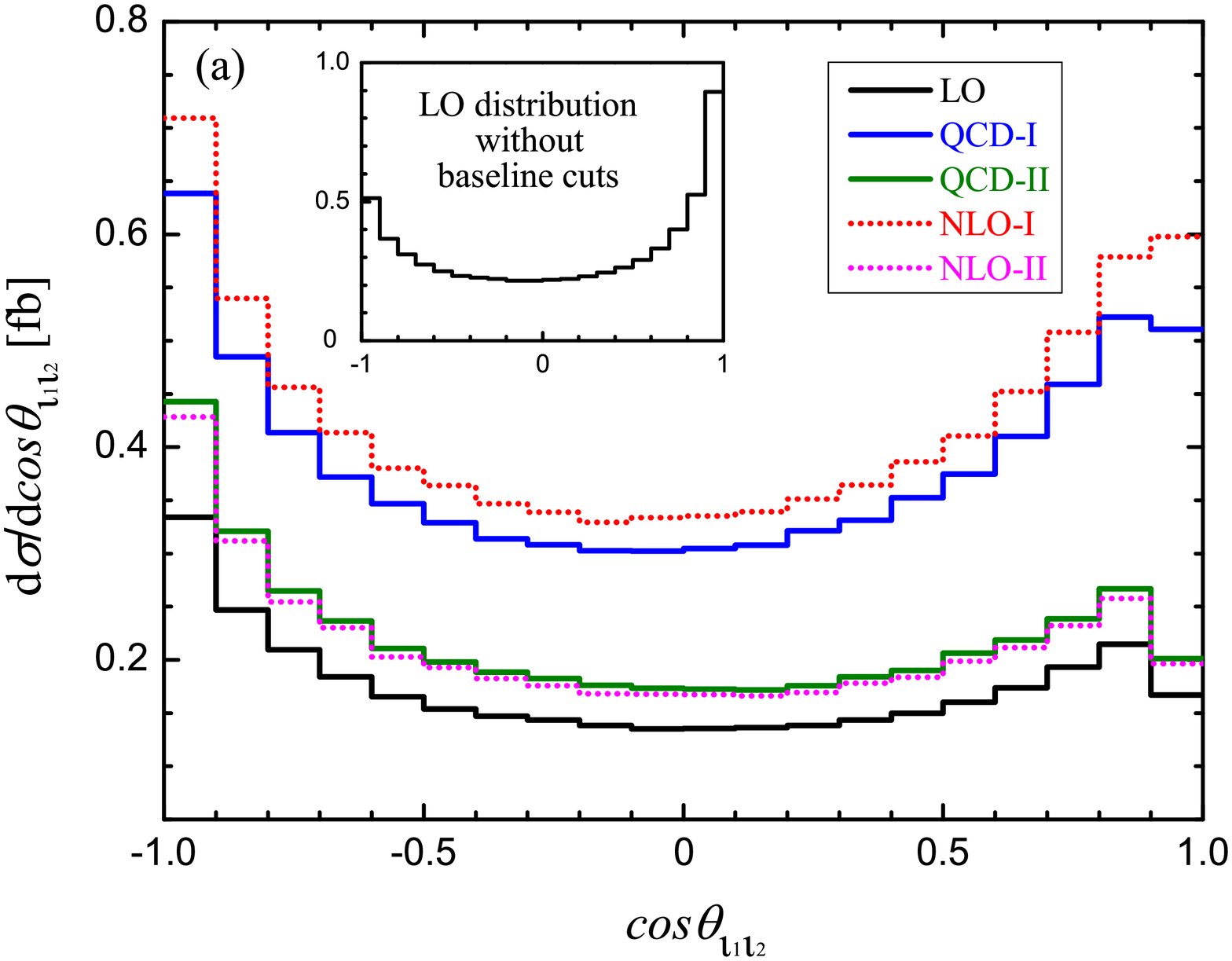}
\includegraphics[width=0.49\textwidth]{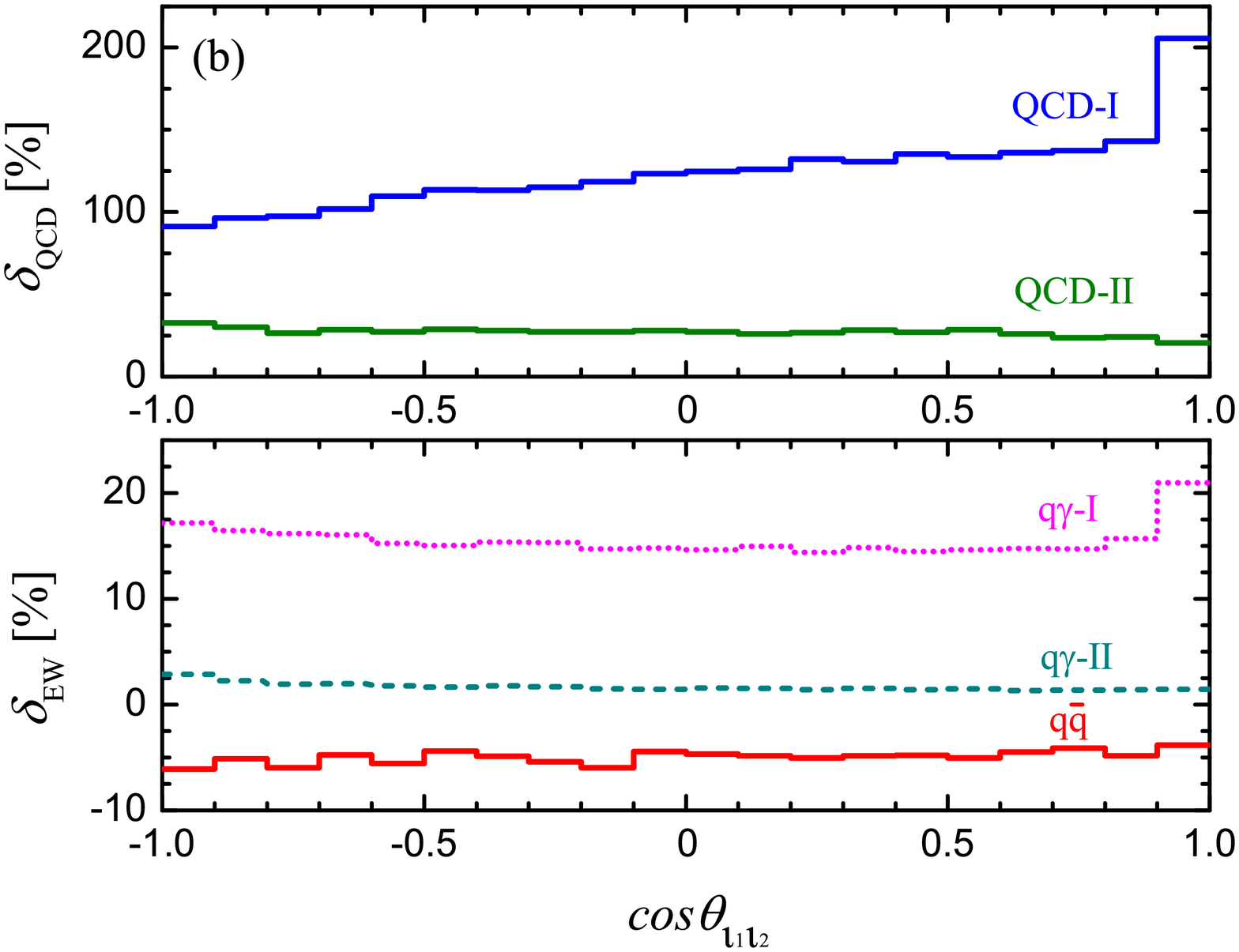}
\caption{
\label{fig-Angle12}
\small
(a) Distributions of the separation angle between the leading and next-to-leading leptons and (b) the corresponding NLO QCD and EW relative corrections for $pp \to W^+W^-W^+ \to 3\ell + 3\nu + X$ at the 14 TeV LHC.
 }
\end{center}
\end{figure}

The LO, NLO QCD, and NLO QCD+EW corrected $\theta_{\ell_1\ell_2}$ distributions for $pp \rightarrow W^+W^-W^+ \rightarrow 3 \ell + 3 \nu + X$ and the corresponding relative corrections are presented in Figs.\ref{fig-Angle12}(a) and \ref{fig-Angle12}(b), separately. The LO $\theta_{\ell_1\ell_2}$ distribution without the baseline requirements of Eq.(\ref{bscuts}) is also depicted in the Fig.\ref{fig-Angle12}(a) inset only for comparison. By comparing the two LO $\theta_{\ell_1\ell_2}$ distributions in Fig.\ref{fig-Angle12}(a), we can see that the leading and next-to-leading leptons in the final state prefer to go out in the same direction, but most of the events in the small $\theta_{\ell_1\ell_2}$ region ($\cos\theta_{\ell_1\ell_2} > 0.9$) are excluded by imposing the baseline cuts at the LO. As shown in Fig.\ref{fig-Angle12}(a), the NLO QCD and QCD+EW corrected distributions in the jet-veto event selection scheme have similar behavior as the LO distribution in the whole $\theta_{\ell_1\ell_2}$ region, while the NLO QCD and QCD+EW corrected distributions in the inclusive event selection scheme exhibit an apparent enhancement in the vicinity of $\cos\theta_{\ell_1\ell_2} \sim 1$ compared to the LO distribution. Correspondingly, Fig.\ref{fig-Angle12}(b) shows that the relative corrections in the jet-veto event selection scheme, $\delta_{\rm QCD}^{\rm (II)}$, $\delta_{\rm EW}^{q\gamma \rm (II)}$, and $\delta_{\rm EW}^{q\bar{q}}$, vary slowly and continuously in the whole $\theta_{\ell_1\ell_2}$ region, while the relative corrections in the inclusive event selection scheme, $\delta_{\rm QCD}^{\rm (I)}$ and $\delta_{\rm EW}^{q\gamma \rm (I)}$, vary smoothly in the region of $\cos\theta_{\ell_1\ell_2} < 0.9$ but increase sharply when $\cos\theta_{\ell_1\ell_2} \rightarrow 1$. It implies that the NLO QCD and EW corrections from the hard jet radiation are hardly reduced by the baseline cuts, and, thus, the considerable enhancement of the relative corrections in the inclusive event selection scheme in the vicinity of $\cos\theta_{\ell_1\ell_2} \sim 1$ is due to the suppression of the LO distribution by the baseline cuts in this region. For example, the QCD and photon-induced EW relative corrections in the inclusive event selection scheme increase slowly from $90\%$ to $145\%$ and vary smoothly around $15\%$ as $\cos\theta_{\ell_1\ell_2}$ increases from $0$ to $0.9$, while they increase sharply to about $205\%$ and $20\%$ in the vicinity of $\cos\theta_{\ell_1\ell_2} \sim 1$, respectively. After applying the jet veto, the QCD and photon-induced EW relative corrections are heavily reduced, particularly in the small $\theta_{\ell_1\ell_2}$ region and are steady at about $25\%$ and $2\%$, respectively, in the whole $\theta_{\ell_1\ell_2}$ region. Since the events with a hard jet are more concentrated in the $\ell_1\ell_2$-collinear region (i.e., $\theta_{\ell_1\ell_2} \rightarrow 0$) and will be excluded by the jet veto, the full NLO QCD+EW relative correction is very sensitive to the event selection scheme in the vicinity of $\cos\theta_{\ell_1\ell_2} \sim 1$.
%--------------------------------------------------------->>>FIG
%---------------------------------------------------------<<<FIG

\par
\begin{figure}[htbp]
\begin{center}
\includegraphics[width=0.49\textwidth]{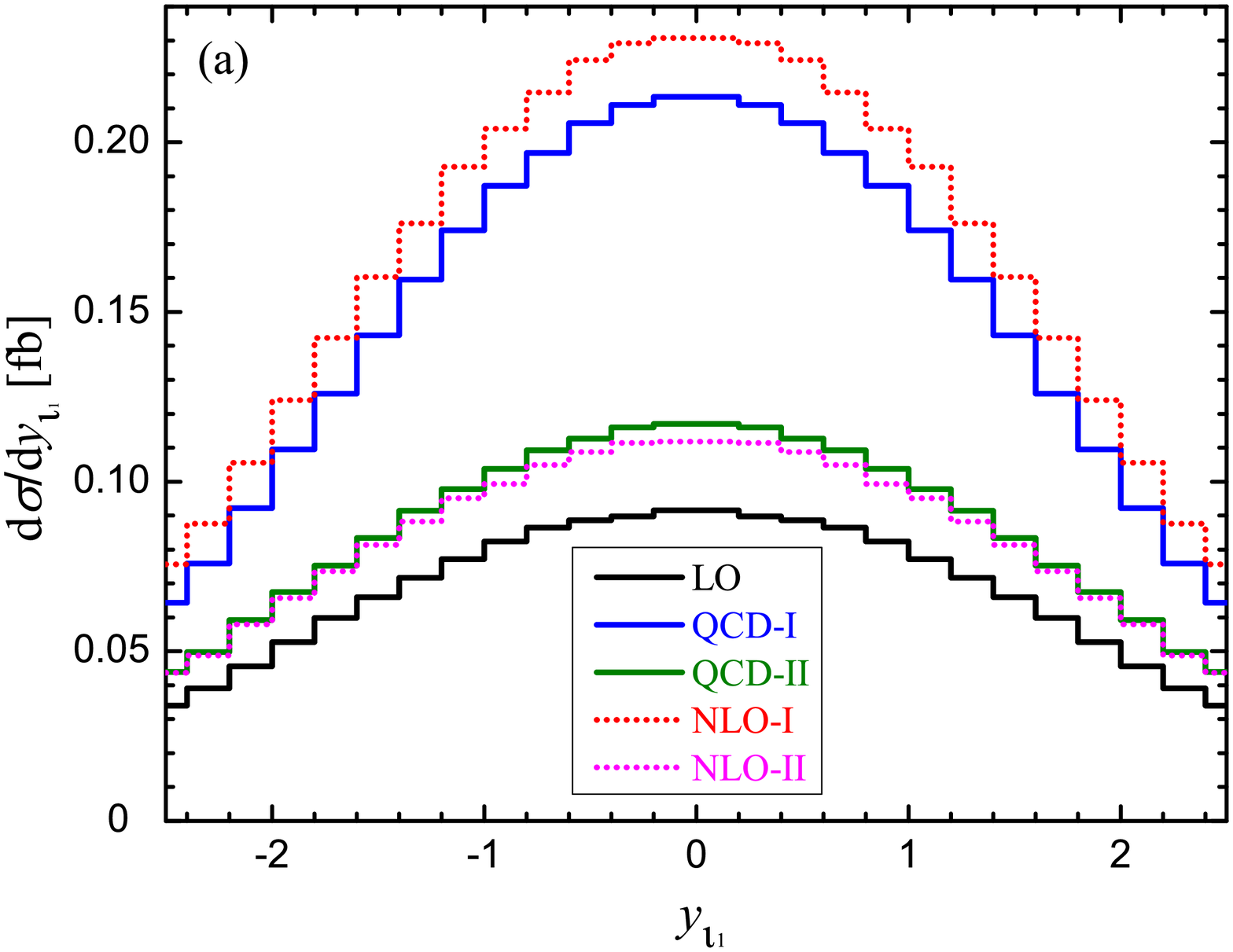}
\includegraphics[width=0.49\textwidth]{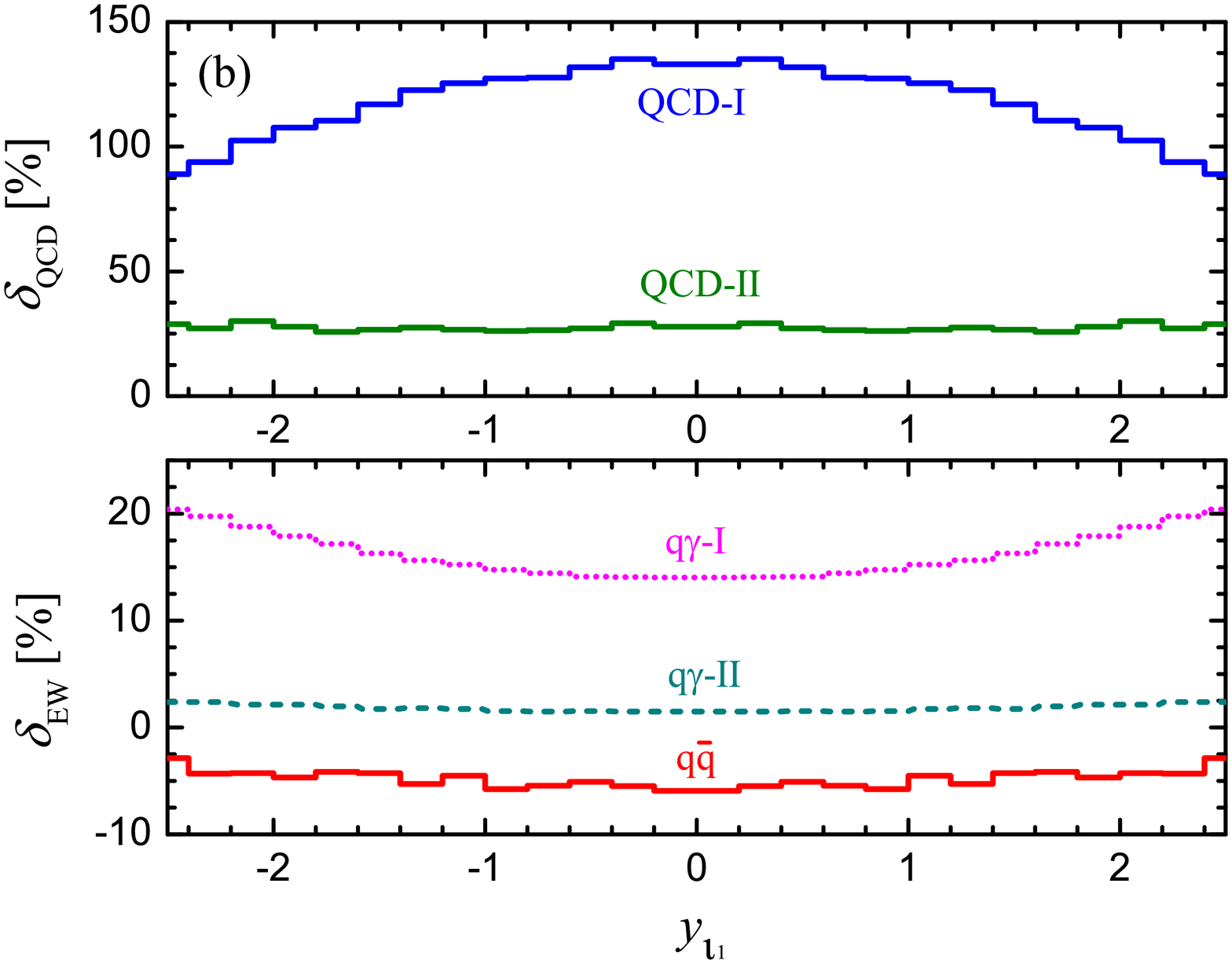}
\caption{
\label{fig-YLl}
\small
(a) Rapidity distributions of the leading lepton and (b) the corresponding NLO QCD and EW relative corrections for $pp \to W^+W^-W^+ \to 3\ell + 3\nu + X$ at the 14 TeV LHC.
 }
\end{center}
\end{figure}

The LO, NLO QCD, and NLO QCD+EW corrected rapidity distributions of the leading lepton and the corresponding relative corrections are given in Figs.\ref{fig-YLl}(a) and \ref{fig-YLl}(b), respectively. In the inclusive event selection scheme, the QCD relative correction approximately decreases from $135\%$ to $90\%$, while the photon-induced EW relative correction increases from $15\%$ to $20\%$ as $|y_{\ell_1}|$ increases from $0$ to $2.5$. In the jet-veto event selection scheme, the QCD and photon-induced EW relative corrections are suppressed and are steady at about $30\%$ and $2\%$, respectively. The EW relative correction from quark-antiquark annihilation is negative, varying in the vicinity of $-5\%$ in the plotted $y_{\ell_1}$ region. Figures \ref{fig-Angle12}(b) and \ref{fig-YLl}(b) clearly show that $\delta_{\rm QCD}^{\rm (II)}$, $\delta_{\rm EW}^{q\gamma \rm (II)}$, $\delta_{\rm EW}^{q\bar{q}}$, and, therefore, $\delta_{\rm NLO}^{\rm (II)}$ are fairly stable in the plotted $\theta_{\ell_1\ell_2}$ and $y_{\ell_1}$ regions. We may conclude that the NLO QCD+EW $K$ factor in the jet-veto event selection scheme is insensitive to the variables of the orientation of final-state leptons.
%--------------------------------------------------------->>>FIG
%---------------------------------------------------------<<<FIG

\vskip 5mm
\section{SUMMARY}

\par
The triple gauge boson production is an ideal platform for determining the triple and quartic gauge couplings and understanding the EW symmetry breaking mechanism as well as the background to new physics beyond the SM. In this work, we calculate the NLO QCD and NLO EW corrections to the triple $W$-boson production with subsequent leptonic decays at the LHC. The {\sc MadSpin} method is employed to take into account the spin correlation and finite-width effects in dealing with the $W$-boson decays. The NLO QCD+EW corrected integrated cross section and some kinematic distributions of final products are provided. Our numerical results show that the NLO QCD correction in the inclusive event selection scheme enhances the LO distributions significantly, particularly in the high-energy region (i.e., the high $M_{WWW}$ or $p_{T, \, \ell_1}$ region) and leads to a comparatively large scale uncertainty due to the QCD real jet radiation. Analogously, the triple $W$-boson events with a hard jet can also be produced via photon-induced channels. The EW correction is sizable in the high $M_{WWW}$ and high $p_{T, \, \ell_1}$ regions. In order to improve the convergence of the perturbative predictions, we employ the jet-veto event selection scheme by vetoing events with a final-state jet of transverse momentum greater than $50~{\rm GeV}$. However, the jet veto will induce an additional theoretical uncertainty which can be improved by adopting the resummation technique.

\vskip 5mm
\section{ACKNOWLEDGMENTS}
This work was supported in part by the National Natural Science Foundation of China (Grants No. 11375171, No. 11405173, No. 11535002, and No. 11375008).

\end{document}